\documentclass[reprint,
aps,
pra
]{revtex4-1}

\usepackage{graphicx}
\usepackage{dcolumn}
\usepackage{bm}
\usepackage[colorlinks=true,linkcolor=blue]{hyperref}

\begin{document}

\preprint{APS/123-QED}

\title{Orientation-dependent dissociative ionization of H$_2$ in strong elliptic laser fields: Modification of the release-time through molecular orientation.}

\author{Arnab Khan}
\email{khan@atom.uni-frankfurt.de}
\author{Daniel Trabert, Sebastian Eckart, Maksim Kunitski, Till Jahnke}
\author{Reinhard D\"{o}rner}
\email{doerner@atom-uni.frankfurt.de}
\affiliation{
 Institut f\"{u}r Kernphysik, J. W. Goethe Universit\"{a}t, Max-von-Laue-Strasse 1, D-60438 Frankfurt am Main, Germany.
}

\date{\today}

\begin{abstract}

We investigate the photoelectron angular emission distributions obtained by strong field dissociative ionization of H$_2$ using cold target recoil ion momentum spectroscopy. In case of employing laser light with an ellipticity close to 0.9 and an intensity of 1.0 $\times$ 10$^{14}$ W/cm$^2$, we find that the most probable release-time of the electron does not generally coincide with the time when the laser field maximizes. The release-time is affected by the molecular orientation. In addition, we observe that the width of the release-time distribution depends on molecular orientation. We attribute this observation to the two-center interference.

\end{abstract}


\maketitle


At which time is an originally bound electron released when an atom is exposed to a strong laser pulse? This question is closely related to the strong field induced tunnel-ionization rates and if such \emph{release-times} depend on the properties of the orbital of an atom, molecules and even solids exposed to the ionizing field \cite{Liu-Kunlong}. A powerful experimental technique to access these ultra-short time scales is attosecond-streaking employing a time-dependent electric field in the THz or optical regime. This streaking field accelerates the electron once it is in the continuum and the momentum transferred depends on the electron release-time. Thus, measuring the final momentum allows one for inferring the release-time of the detected electron with attosecond precision \cite{Itatani_Atto,corkum2007attosecond}. A particularly intuitive implementation of this idea is angular streaking in which one uses close to circularly polarized femtosecond pulses \cite{eckle2008attosecond,Eckle_Science,wu2012probing,wu2013understanding,landsman2014ultrafast,sainadh2019attosecond}. In strong field ionization employing circularly
polarized light the emission direction of the electron (in the polarization plane) encodes the release-time of the electron with high precision: The rotating electric field of the femtosecond laser acts as an ultrafast clockwork and the measured electron momentum vector serves as the hand of the clock. This concept is well-known as the `attoclock' \cite{Gallmann_attoclock}. Despite its conceptual simplicity, this technique paved the way to measure attosecond phenomena with femtosecond laser pulse. The most widely used observable in angular streaking experiments is the angle at which the electron count rate maximizes. After accounting for the influence of the ionic Coulomb field, this angle is usually associated with the time at which the electron most likely appears in the continuum (i.e. the `release-time'). A second important observable is the width of this angular distribution. Until today angular streaking or attoclock measurements are mostly limited to atomic targets \cite{eckle2008attosecond,Eckle_Science} and only very few cases involving molecules or molecule-like species have been reported so far \cite{wu2012probing,wu2013understanding,QuanPRL}. If an atom is ionized by a strong laser field, the photoelectron most likely escapes the atomic potential at the peak of the laser's electric field, because the tunneling process depends in a strongly nonlinear manner on the laser intensity. For a molecule, the ionization rate depends additionally on the orientation of the molecular axis with respect to the instantaneous electric field vector \cite{PhysRevA.66.033402}, which has been attributed to the shape of the ionized molecular orbital.

\begin{figure}[htb!]
\includegraphics[scale=1]{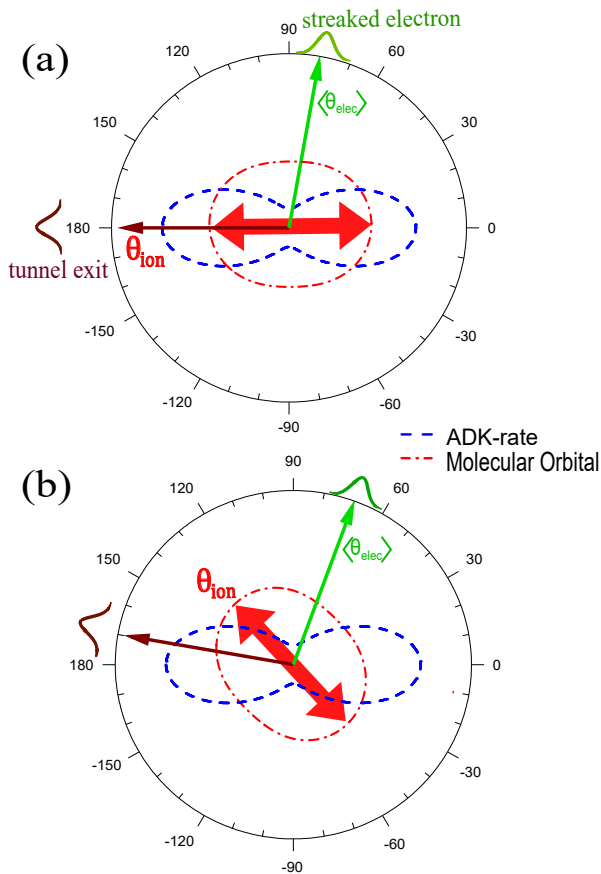}
\caption{\label{fig:1} (Color online) Schematic representation of the molecular attoclock-scheme for a clockwise rotating laser field. The scheme allows for a mapping of the electron release-time to momentum space that is experimentally accessible. The two-sided red (dark gray) thick arrow corresponds to the molecular axis. The brown (black) thin arrow represents the direction of the tunnel exit in the plane of polarization at the instant of ionization (release-time). The green (light gray) thin arrow indicates the most probable electron emission-angle ($\langle\theta_{elec}\rangle$) which corresponds to this release time. (a) The molecular axis is oriented along the peak of the laser electric field.  (b) Same as (a) but for a different molecular orientation. The ionization probability differs due to the changed relative angle of the tunneling direction and the molecular orientation [red (dark gray) dash-dotted line]. As a result, the release-time is altered which yields a change of the most probable electron emission-angle ($\langle\theta_{elec}\rangle$). All the angles are measured relative to the major axis of the polarization ellipse.}
\end{figure}

In this paper, we investigate how the interplay between the molecular orbital and the driving laser field influence the polarization plane photoelectron angular emission distribution in dissociative ionization of H$_2$. We show, that in case of strong-field ionization of H$_2$, the most probable electron emission angle (and thus the most probable release-time) depends on the molecular orientation with respect to the laser polarization. Additionally, it is demonstrated that the width of this angular distribution depends on the molecular orientation, too. The latter issue has been recently addressed theoretically by Serov \textit{et al.} in their numerical simulation of the attoclock approach \cite{Serov2019PRA}. Fig. \ref{fig:1} illustrates the scheme which we employed to access the electron release-time as a function of the molecular orientation.

In this experiment, elliptically polarized intense laser pulses with a duration of 100 fs at a central wavelength of 790 nm were generated by a Ti:sapphire femtosecond laser system (Wyvern-500, KMLabs). The ellipticity ($\epsilon$) of the pulse was approximately 0.9. To produce elliptically polarized pulses, we used a combination of a quarter-wave ($\lambda/4$) plate and a half-wave ($\lambda/2$) plate which were placed before the entrance window of the experimental chamber. Inside the vacuum chamber, the laser beam was focused by a spherical concave mirror ($f$=60 mm) onto a cold supersonic H$_2$ gas-jet. The gas-jet was produced by means of a supersonic  expansion of H$_2$ gas through a 30 $\mu$m nozzle with a driving pressure of 1.2 bar at room temperature (300 K). We measured the three-dimensional momentum distributions of the H$^+$-ion and the electron, in coincidence, by using the cold target recoil ion momentum spectroscopy (COLTRIMS) technique \cite{DORNER200095,Ullrich}. The spectrometer used here has an ion arm consisting of a 18.2 cm acceleration region and a 40.0 cm drift region and an electron arm with an acceleration length of 7.8 cm. The  H$^+$ ions and photoelectrons were accelerated by a homogeneous electric field of 42.3 V/cm towards two micro-channel plate detectors equipped with delay-line anodes \cite{Jagutzki2002}. A superimposed magnetic field of 10.1 Gauss was employed in order to confine the electrons inside the spectrometer volume. We used a laser intensity of 1.0 $\times$ 10$^{14}$ W/cm$^2$. The intensity calibration was done by examining the photoelectron distribution from single ionization of argon using a circularly polarized laser pulses. To obtain an accurate measure of the peak intensity at the center of the focal volume, the experimental electron radial momentum ($p_{er}$=$\sqrt{p_{ey}^2 + p_{ez}^2}$) distribution was compared with Monte-Carlo based semi-classical two-step (SCTS) model \cite{SCTS}. In this simulation non-adiabatic initial conditions were extracted from the Strong field approximation (SFA) \cite{Eckart_Banana}. Using this method, we obtained a peak electric field strength of 0.043 a.u. for the elliptically polarized laser pulse.

Strong field dissociative-ionization of H$_2$ is typically considered a two-step process:

\begin{equation}\label{eq:1}
 \mathrm{H_2} + n\mathrm{\hbar\omega} \rightarrow \mathrm{H_2{}^+} + {e^-}  
\end{equation}
\begin{equation}\label{eq:2}
 \mathrm{H_2{}^+} + m\mathrm{\hbar\omega} \rightarrow \mathrm{H^+} + \mathrm{H^0} 
\end{equation}
In the very first step, the H$_2$ molecule is photo ionized (see Eq. \ref{eq:1}). The resulting H$_2{}^+$ ion is created in a broad superposition of vibrational states (on $1s\sigma_g^+$ potential curve) and further nuclear dynamics is driven by the laser field. During its vibational motion the H$_2{}^+$ ion can absorb further photons from the laser field and dissociate into H$^+$ and H$^0$ (see Eq. \ref{eq:2}) \cite{PhysRevLett.103.213003,PhysRevLett.103.223201,PhysRevLett.108.063002,wu2013understanding}. There are two well-known channels for H$_2{}^+$ dissociation: The 1-photon (1$\omega$) channel and the net-2-photon (net-2$\omega$) channel. In the 1$\omega$ channel, the H$_2{}^+$ ion absorbs one photon from the laser field and is electronically excited to the repulsive $2p\sigma_u^+$ potential energy curve on which it dissociates. In the net-2$\omega$ channel, the H$_2{}^+$ ion is excited to the $2p\sigma_u^+$ potential energy curve by absorbing three photons at small internuclear distance. The nuclear wave-packet generated moves along that potential energy-curve to a larger internuclear distance at which one photon is emitted, yielding a transition to the $1s\sigma_g^+$ potential curve along which the molecule finally dissociates. One can easily separate these two channels experimentally by the proton kinetic energy spectra (or proton momentum-distribution in the laser polarization plane). The net-2$\omega$ channel yields higher kinetic energy than the 1$\omega$ dissociation.  
Here, we discuss only on the net-2$\omega$ channel as this channel is easy to distinguish regarding inter-channel mixing compared to the 1$\omega$ channel \cite{Sturm_H2}.

\begin{figure}[htb!]
\begin{centering}
\includegraphics[scale=1]{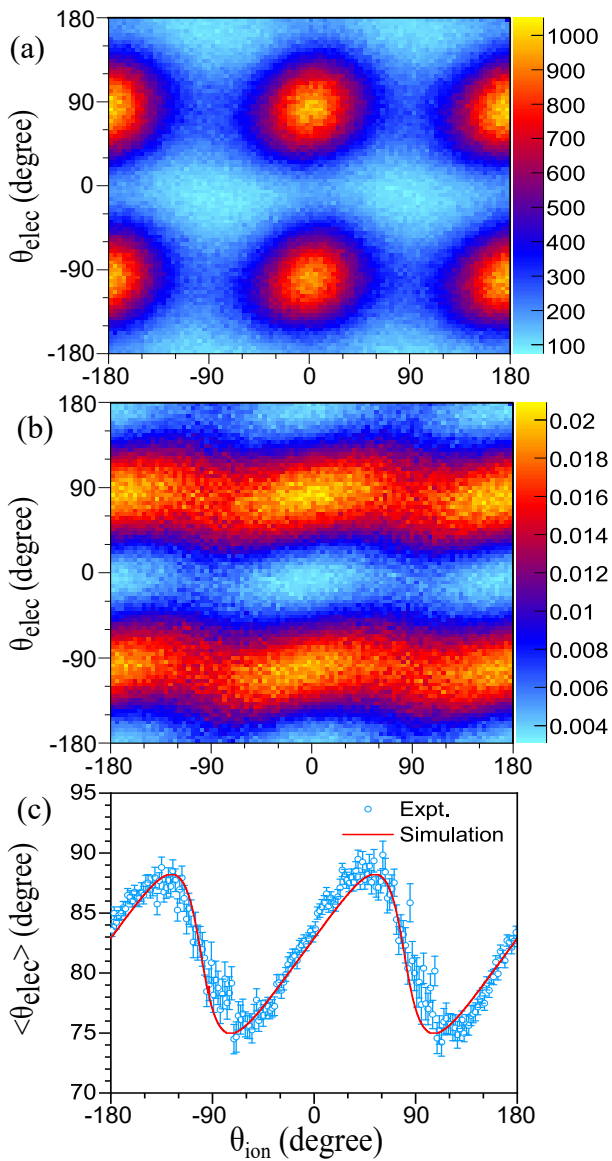}
\caption{\label{fig:2}(Color online) (a) Correlation plot of electron emission-angle ($\theta_{elec}$) and the ion emission-angle ($\theta_{ion}$) within the polarization plane. The plot is restricted to the net-2$\omega$ dissociation channel. 
(b) Same plot as (a) but after normalizing each column to one (see text).  
(c) Blue (gray) circles profile of (a) depicting the most probable electron emission-angle ($\langle\theta_{elec}\rangle$). The profile has been created by restricting to an angular range of 0$^\circ$ $<$ $\theta_{elec}$ $<$ 180$^\circ$ [in Fig. \ref{eq:2}(a)]. The peak of the electron angular distribution occurs at +81.0$^\circ$ in this range. (See Fig. \ref{fig:1} for the definition of the angles.) Red (dark gray) solid line: Corresponding profile obtained from our simulation.}
\end{centering}
\end{figure}

\begin{figure*}[htb!]
\begin{centering}
\includegraphics[scale=1]{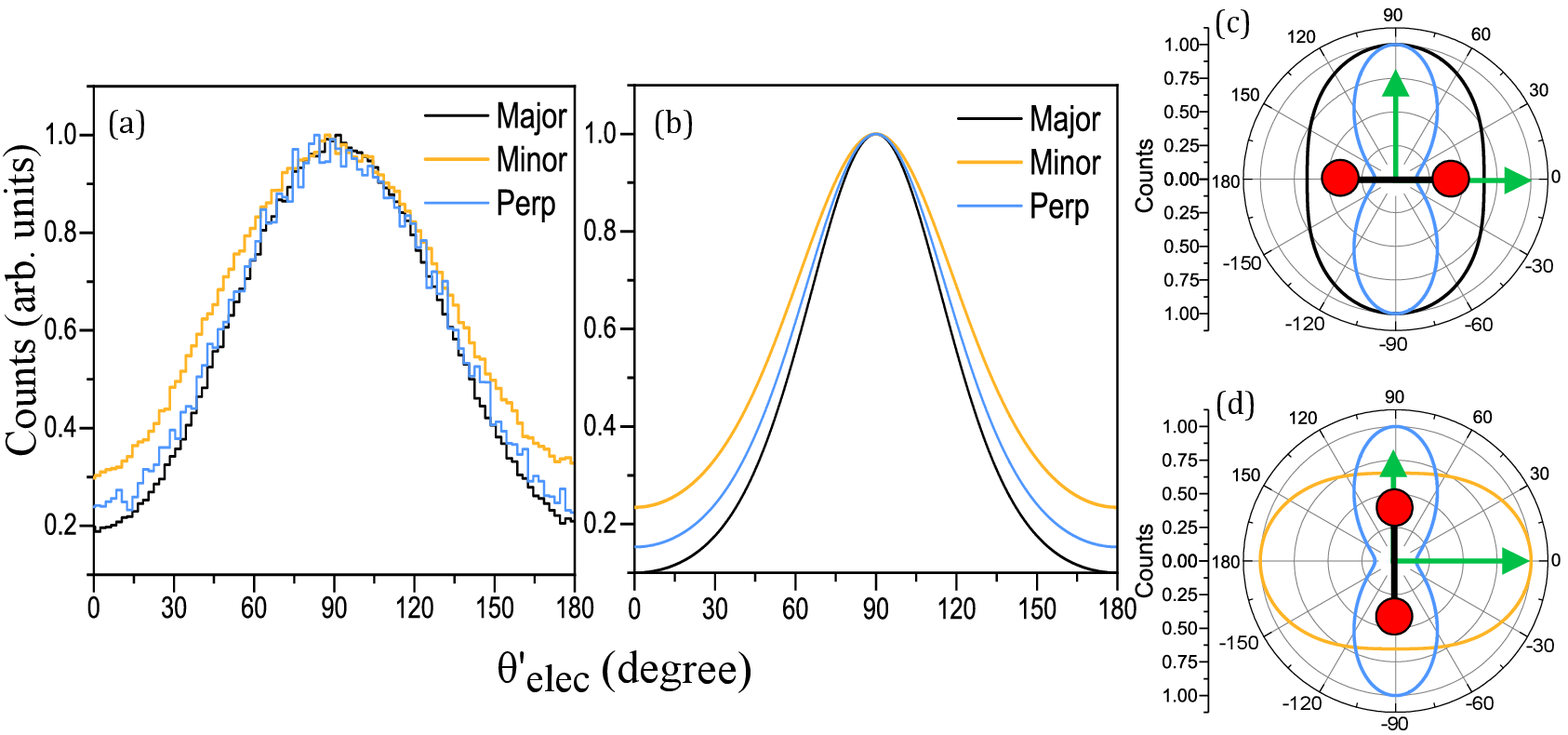}
\caption{\label{fig:3} (Color online) Photoelectron angular distribution in the polarization plane for three different orientations of the H$_2$ molecule with respect to the laser polarization: (a) Experimental distributions ($\theta^{'}_{elec}$ = $\theta_{elec}$ + 9.0$^\circ$: the experimental distribution has been shifted to 90.0$^\circ$ for an easy comparison with the simulation), (b) Calculated distribution incorporating the interference pattern ($\mathbf{P_{TC}}$ for $k_{elec} = 0.9$ a.u.) using Eq. \ref{eq:3} modulated by $\mathbf{P_{SC}}$($\theta_{elec})$. The two panels on the right depict the interference pattern ($\mathbf{P_{TC}}$, broad distributions) for different orientations of the molecule [given by the red (dark gray) circles connected by the black solid line] along with the $\mathbf{P_{SC}}$($\theta_{elec})$ (narrow distributions always aligned along 90.0$^\circ$ and -90.0$^\circ$). The arrows show the direction of the major and minor axis of the laser field. (c) The H$_2$ molecule is aligned with the major axis, (d) with the minor axis. }
\end{centering}
\end{figure*}

Fig. \ref{fig:2}(a) shows the measured correlation of the electron emission-angle ($\theta_{elec}$) and the ion emission-angle ($\theta_{ion}$)  in the polarization plane (see Fig. \ref{fig:1} for the definition of the angles). After performing a column-wise normalization of the correlation plot [see Fig. \ref{fig:2}(b)], it is evident that the distribution has a varying slope, and it is not equal to one. Hence, for a particular variation of $\theta_{ion}$, $\theta_{elec}$ varies less. For example, in the range -70$^\circ$ $<$ $\theta_{ion}$ $<$ 50$^\circ$ a change of $\theta_{ion}$ of almost 110$^\circ$ results in a change of only 15$^\circ$ in $\theta_{elec}$. To visualize this, the most probable electron emission-angle $\langle\theta_{elec}\rangle$  is plotted as a function of $\theta_{ion}$ in Fig. \ref{fig:2}(c). This observation can be understood considering a simple model in which the ionization probability is a product of the photoelectron angular distribution for perfectly circularly polarized light which mimics the shape of the molecular orbital and the Ammosov-Delone-Krainov (ADK) rate \cite{Brichta_2006}. Since the variation in ADK-rate (which is a property of the laser field) is much steeper than the molecular orbital angular distribution (which is a property of the molecule), the electron angular emission distribution is dominated by the properties of the laser electric field while the molecular orientation has only a week effect. This is illustrated in Fig. \ref{fig:1} and the quantitative result from the simple model is shown in Fig. \ref{fig:2}(c) [red (dark gray) solid line]. 

For this quantitative modelling, we use a parametrization of the measured electron angular distribution for circularly polarized light  $\sqrt{\sin^2\theta + \eta^2\cos^2\theta}$ \cite{PhysRevLett.102.033004}; Here, $\eta$ reflects the asymmetry between the long and short axis of the angular dependent ionization probability. We find a value of $\eta = 1.55$ which has been optimized to match the experimental variation of the angular profile [see Fig. \ref{fig:2}(b) and (c)]. This value matches perfectly with the molecular strong-field approximation (MOSFA) calculation as reported in Ref. \cite{PhysRevLett.102.033004}.

Neglecting Coulomb interaction after tunneling, the electron emission-angle ($\theta_{elec}$) is determined by the vector potential at the instant the electron exits the tunnel (i.e. the release-time). We assumed that Coulomb interaction after tunneling introduces an offset that is independent of the release-time of the electron. \cite{PhysRevA.83.023405}. Thus, the Coulomb interaction after tunneling leads to a constant shift in the streaking angle (see Fig. \ref{fig:1}). A classical model calculation further confirms this negligible change in the offset angle (on the level of 0.1$^\circ$) for different orientations of H$_2$. Within this assumption, our measurement provides direct access to the photoelectron release-time difference for different molecular orientations. The observed variation of the streaking angle $\langle\theta_{elec}\rangle$  in [Fig. \ref{fig:2}(c)] of approximately 15$^{\circ}$ $\pm$ 3$^{\circ}$ corresponds to a difference in the most probable release-time of 110 $\pm$ 22 attoseconds.

In a next step, we explore how the width of the photoelectron angular distribution depends on the molecular orientation with respect to the major axis of the polarization ellipse of the laser pulse. In Fig. \ref{fig:3}(a) we show the distribution of $\theta_{elec}$  in the plane of polarization for three different orientations of H$_2$ (along the major-axis, minor-axis and perpendicular to the polarization plane). If the molecule is oriented along the minor-axis of the light's polarization ellipse, the photoelectron distribution is broader; while for the major-axis alignment case, the distribution is narrower. The width of the angular distribution for the perpendicular orientation is in-between the two other cases. The three orientations correspond to cases where the H$^+$-ions are ejected into an angular interval of $\pm$15$^{\circ}$ around the maximum (major-axis) and minimum (minor-axis). For the third case we select H$_2{}^+$ ions emitted into an angular interval of $\pm$40$^{\circ}$ with respect to the light propagation direction.

We follow Ref. \cite{Serov2019PRA} and argue that this change of width of the emission pattern is a consequence of two slit interference.  For its fundamental nature, H$_2$ has long been a strong candidate for investigating the impact of two-center interference on many interaction processes like ion or electron scattering and electron emission induced by ion, electron and photon impact \cite{Schmidt_ion_scat,Alexander_ion_scat,Deepankar_e_emission_ion,electron_emission_e2e,Akoury949,PhysRevA.78.013414,PhysRevLett.117.083002}. in case of laser-induced ionization, the large tunnel-exit position and the Coulomb interaction with the residual molecular ion make the double slit analogy less obvious \cite{kunitski2019double}. However, very recently laser field-induced double-slit interference has been seen by Kunitski \textit{et al.} in the molecular frame photoelectron angular distribution of dissociating Ne$_2{}^+$ ions \cite{kunitski2019double}. To estimate the effect of the two-center interference on the width, we approximate the angular distribution as a product of an angular distribution from a single center emission [$\mathbf{P_{SC}}$($\theta_{elec})$] and an angular distribution $\mathbf{P_{TC}}$ given by the interference between the emisssion contributions from two single centers separated by the internuclear distance $R$ with an electron momentum $k_{elec}$ as suggested by \cite{Serov2019PRA}. For an orientation of the internuclear axis in the polarization plane at an angle $\theta_{ion}$ we obtain:

\begin{equation}\label{eq:3}
\mathbf{P_{TC}} \propto \cos^2\left(k_{elec}\frac{R}{2} \cos(\theta_{elec}-\theta_{ion})\right)
\end{equation}

The double slit contribution $\mathbf{P_{TC}}$ is shown in Figs. \ref{fig:3}(c) and (d) for the two cases of the molecule being aligned along the major  ($\theta_{ion}$ = 0$^\circ$) and minor ($\theta_{ion}$ = 90$^\circ$) axis of the polarization ellipse.
The single center angular distribution $\mathbf{P_{SC}}$($\theta_{elec})$ one can roughly estimate using the ADK rate for an ionization potential of $I_p = 15.4$ eV and the field strength as given by the polarization ellipse. The theoretical result of $\mathbf{P_{TC}}$ and the product  $\mathbf{P_{SC}}$($\theta_{elec})$ $\cdot$ $\mathbf{P_{TC}}$ is shown in Fig. \ref{fig:3}(b). 

This simple model nicely reproduces the experimental observation of the variation of the width of the electron angular distribution with molecular orientation [see in Fig. \ref{fig:3}(a)]. The physical effect is, that for an orientation of the molecule parallel to the major axis the zero order interference fringe provides a function which peaks along the maximum of the single center distribution thus narrows its down, while for the molecular orientation perpendicular the central fringe of the double slit points at $\theta_{elec}$= 0$^\circ$, 180$^\circ$ [Fig. \ref{fig:3}(c)] enhancing the wings of the angular distribution.

In conclusion, in this paper, we demonstrate that the interplay between the shape of a molecular orbital and the ADK-rate allows to intuitively understand attosecond electron release-time in dissociative ionization of H$_2$ using the angular streaking method. In contrast to the atomic case, we can disentangle the electron release-time difference occurring for different molecular orientations with respect to the laser polarization vector. This has been done by analyzing the photoelectron and ion angular emission distributions measured in coincidence.

\begin{acknowledgments}
This work is supported by Deutsche Forschungsgemeinschaft (DFG). We also acknowledge partial support from Deutsche Forschungsgemeinschaft via Sonderforschungsbereich 1319 (ELCH). AK thanks Alexander von Humboldt foundation for the support.
\end{acknowledgments}


\providecommand{\noopsort}[1]{}\providecommand{\singleletter}[1]{#1}%

\end{document}